\documentclass[twocolumn,showpacs,preprintnumbers,amsmath,amssymb]{revtex4}

\usepackage{graphicx}
\usepackage{dcolumn}
\usepackage{bm}

\begin{document}


\title{Hybrid (kinetic-fluid) simulation scheme based on method of characteristics}

\author{N. Javaheri}
 \author{S. Rahimi}
 \author{H. Abbasi}
\email{abbasi@aut.ac.ir}
 \affiliation{Faculty of Energy Engineering and Physics, Amirkabir University of Technology, P. O. Box\\
15875-4413, Tehran, Iran}
\date{\today}

\begin{abstract}
Certain features of the method of characteristics are of considerable interest in relation with Vlasov simulation [H. Abbasi {\it et al}, Phys. Rev. E \textbf{84}, 036702 (2011)]. A Vlasov simulation scheme of this kind can be recurrence free providing initial phase points in velocity space are set randomly. Naturally, less filtering of fine-structures (arising from grid spacing)  is possible as there is now a smaller scale than the grid spacing that is average distance between two phase points. Its interpolation scheme is very simple in form and carried out with less operations. In our previous report, the simplest model (immobile ions) was considered to merely demonstrate the important features. Now, a hybrid model is introduced that solves the coupled Vlasov-Fluid-Poisson system self-consistently. A possible application of the code is the study of ion-acoustic (IA) soliton attributes. To this end, a collisionless plasma with hot electrons and cold positive ions is considered. For electrons, the collisionless Vlasov equation is solved by following collisionless phase point trajectories in phase space while ions obey the fluid equations. The periodic boundary conditions are assumed. Both, the characteristic equations of the Vlasov equation and the fluid equations are solved using the Leapfrog-Trapezoidal method. However, to obtain the first half-time step of the Leapfrog, the Euler-Trapezoidal scheme, is employed. The presented scheme conveniently couples the two well-known grids in the Leapfrog method. The first test of the model is an stationary IA soliton. Trapping of electrons is considered and the associated phase space hole is shown. Then as a non-stationary test, the IA soliton generation from a localized initial profile is examined. Conservation laws are the other benchmark tests.             
\end{abstract}

\pacs{47.11.-j, 05.10.-a,  52.30.-q, 52.65.-y}

\maketitle

\newpage
\section{Introduction}
In some plasma devices, such as the Q-machine or plasma discharge, ions temperature is at least one order of magnitude less than electrons temperature. In such plasma devices,  Therefore, thermal effects, associated with the ions, are negligible. Accordingly, for the study of phenomena involving kinetic effects (such as electron trapping), one deals with solving the Vlasov equation for the electrons together with the fluid equations (the continuity and momentum) for the ions. An example of this situation is the generation of ion-acoustic (IA) soliton due to the nonlinear decay of a localized perturbation \cite{ionacoustic1,ionacoustic2,ionacoustic3,ionacoustic4,Lonngren, Hakimi}. 

Numerical simulations of the Vlasov equation are of fundamental importance for the study of many nonlinear processes in kinetic plasmas. The knowledge of the temporal evolution of the distribution function has long been a desire of plasma physicists as well as many involved in many-body physics researches. Particle trapping is an example where the temporal evolution of particle distribution function has to be considered. Many researchers have done great efforts with some success in the numerical integration of the Vlasov equation
(see the Refs. \cite{Vlasov1,Vlasov2,Vlasov3,Vlasov4,1,2,3} and references therein).

In the present work, an unbounded collisionless plasma composed of the cold positively charged ions and hot electrons is considered under the electrostatic approximation. The Vlasov equation is solved for the electrons. We directly follow the characteristics along which the distribution function of electrons, is constant \cite{2}. We choose some representative phase points, in the phase space, that are initialized by an initial distribution function. The phase points following the characteristics are advanced in time by a predictor-corrector method. Interpolation is performed between the phase points and a fixed grid in the phase space to obtain the distribution function on the grid. From the latter, all the desirable quantities, such as the electron charge density, is obtained and used in the Poisson equation. Since the ions are assumed to be cold, their dynamics are governed by the fluid equations. Matching of the two different simulation schemes, needs great attention and is the main motivation behind this paper. First test of the hybrid code is about propagation of a stationary IA soliton. Electron trapping has been considered in the test problem as the result of their nonlinear resonant interaction with the IA soliton. As a non-stationary experiment, soliton generation from an initial Gaussian profile is considered. The conservation of total energy and entropy would be the other benchmarks.  In order to avoid the error, associated with the periodic boundary condition, another version of the code, based on moving grid, is introduced. 

The paper is organized as follows. Section II deals with the basic equations. Section III is devoted to the details of hybrid algorithm. Section IV is briefly devoted to first the calculation of a stationary IA soliton as a test problem and then introducing a non-stationary test problem, i.e. disintegration of a Gaussian profile into IA solitons. The results of the code performance is presented in Sec. V. The paper terminates in section VI by a brief conclusion.

\section{Basic equations}

Let us consider, the one-dimensional electrostatic system, governing on the plasma dynamics with characteristic frequency close to the plasma frequency of ions. Therefore, the ion dynamics is of great importance. However, we assume the thermal effects associated with the ions is negligible. Thus, the fluid equations are convenient for the ionic part of the dynamics. The electrons are treated kinetically. That is, the Vlasov equation governs the electronic part. Poisson equation is the closure. In summary, we have the following set of equations:
\begin{eqnarray}
&& \frac{\partial f_e}{\partial t} + v_e \frac{\partial f_e}{\partial x} + \frac{1}{\alpha}\frac{\partial \phi}{\partial x}\frac{\partial f_e}{\partial v_e} = 0, \label{1}\\ 
&& n_e = \int f_e dv_e, \label{2} \\ 
&& \frac{\partial n_i}{\partial t} + \frac{\partial }{\partial x}\left(n_i v_i\right) = 0, \label{3}\\ 
&& \frac{\partial v_i}{\partial t} + \frac{\partial }{\partial x} \left(\frac12 v_i^2 + \phi \right) = 0, \label{4}\\ 
&& \frac{\partial^2 \phi}{\partial x^2} = n_e -n_i, \label{5}
\end{eqnarray}
where $f_e$ is the electron distribution function, $v_e$ is the electron phase space velocity, $\phi$ is the self-consistent electric potential, $\alpha$ is the ratio of electron mass to ion one ($=m_e/m_i$), $n_e$ ($n_i$) is the electron (ion) density, $v_i$ is the ion fluid velocity, and the following normalizations are used, 
\begin{eqnarray}
&&\omega_{pi} t \equiv t,~~~\frac{x}{\lambda_D}\equiv x,~~~\frac{n_e}{n_0}\equiv n_e,~~~\frac{n_i}{n_0}\equiv n_i,\nonumber\\
&&\frac{c_s f_e}{n_0}\equiv f_e,~~~\frac{v_e}{c_s}\equiv v_e,~~~\frac{v_i}{c_s}\equiv v_i,~~~\frac{e\phi}{T_e}\equiv \phi.
\end{eqnarray}
In the above, $\omega_{pi}=(4\pi e^2n_0/m_i)^{1/2}$, $\lambda_D=[T_e/(4 \pi e^2n_0)]^{1/2}$, $n_0$ is the equilibrium value of particle densities when there is no plasma perturbation, $c_s=(T_e/m_i)^{1/2}$, $e$ is the magnitude of the electron charge, and $T_e$ is the electron temperature. 

\section{The model}

As it was mentioned, the electrons dynamics is the kinetic part of this hybrid code and is governed by the Vlasov equation [Eq. (\ref{1})]. In order to solve it, we directly follow the characteristics along which $f_e$ is constant.
\begin{figure}
\includegraphics[height=7cm,width=9cm]{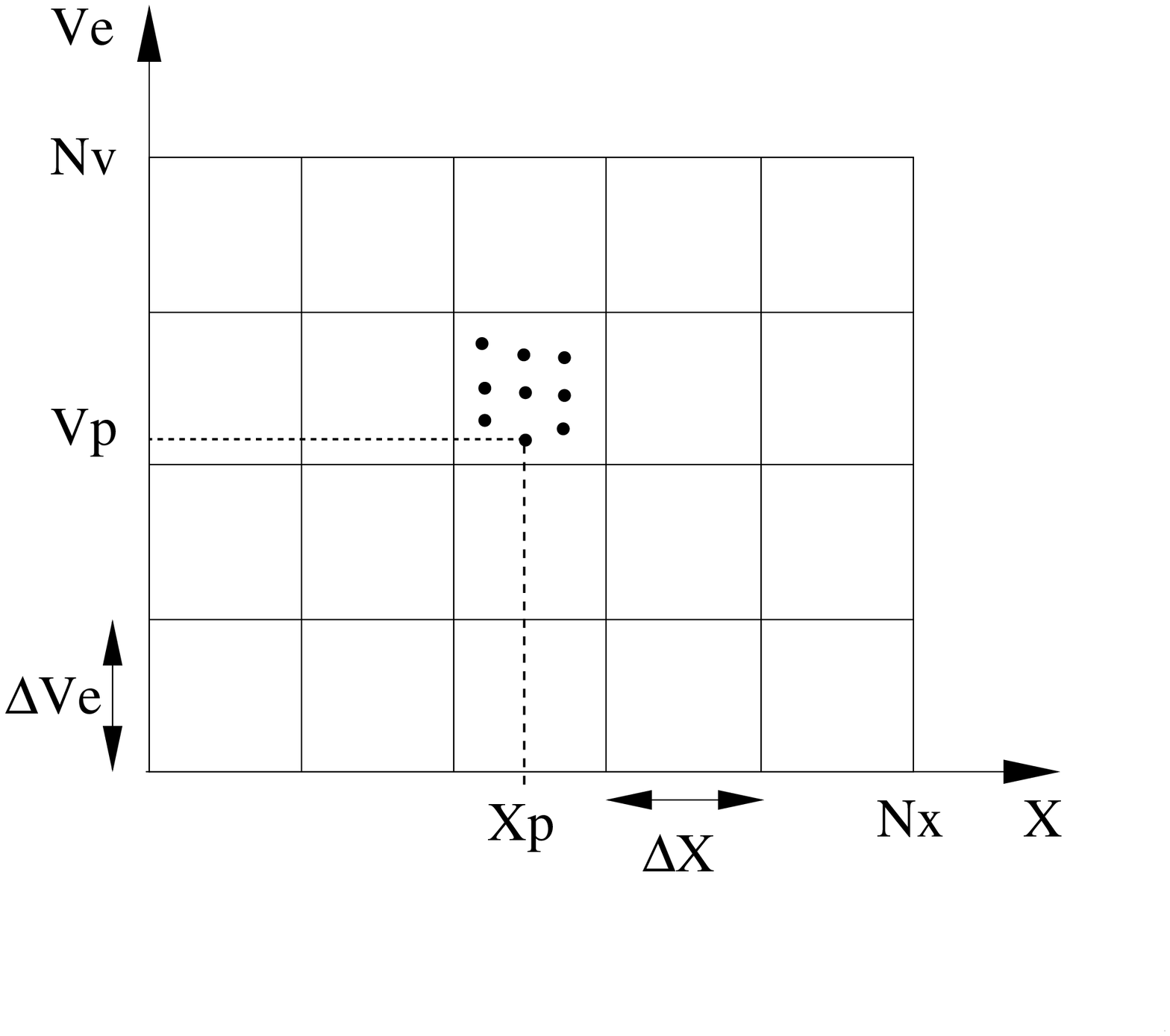} \\
\caption{\label{fig1}A typical phase space grid with $Nx\times Nv$ grid points and 9 phase points
in a host cell. $\Delta x$ and $\Delta v_e$ correspond to the respective grid
spacings.}
\end{figure}

Figure 1 exhibits a typical fixed grid with $Nx \times Nv$ grid points (the vertices of rectangulars) and nine phase points (black circles) in a cell. Thus, from hereon, we introduce the subscript ``$p$" and ``g" to denote the quantity at the phase point and grid point positions in the phase space, respectively. As it was mentioned in Ref. \cite{2}, the accuracy of the code depends directly on the number of phase points. Figure 2 depicts the relative error in the total energy for three different cases. The curves are the results of the IA soliton experiment which is explained in Sec. V. Obviously, for larger population of phase points, the accuracy of the code is enhanced.  Accordingly, we put nine phase points in each cell that is initially set regularly along $X$ axis and to prevent the recurrence effect, randomly along $V_e$ axis (Fig. 1). 
\begin{figure}
\includegraphics[height=7cm,width=9cm]{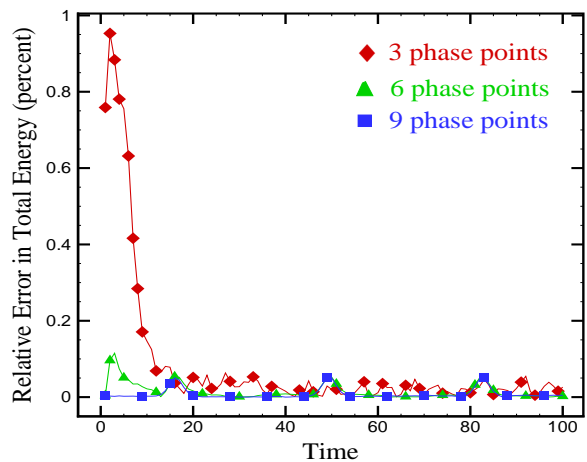} \\
\caption{\label{fig2}The relative error in the total energy for three different population of phase points.}
\end{figure}
Each phase point is by definition characterized by its position $x_p$ and its velocity $v_p$, and has associated with it a distribution function value $f_p$. As the phase points follow their collisionless trajectories, according to the following characteristic equations of Eq. (\ref{1}), 
\begin{eqnarray}
&&\frac{dx_p}{dt} = v_p, \label{7}\\
&&\frac{dv_p}{dt}=-\frac{E_p}{\alpha}, \label{8} 
\end{eqnarray}
$f_p$  remains unchanged, where $E_p$ is the electric field at the phase point position. As the representative phase points follow their characteristics, they continually exchange information with the fixed background grid. Each representative phase point contributes its distribution function to the corners of its instantaneous host cell. In this way, the grid distribution function, $f_g$, is calculated from $f_p$ by the "average interpolation scheme" introduced in Ref. \cite{2}. 

\subsection{The initial loop of the code} 

Let us specify each quantity at time $t=n\Delta t$ by a superscript ``n". After setting the phase points on the phase space and allocating to each of them a distribution function value, we know $x_p^0$, $v_p^0$, and $f_p$ for the Vlasov part of simulation. For calculating the initial value of the electric potential, we proceed as follows. First, by the interpolation $f_g^0$ is computed from $f_p$.  Then, by integrating $f_g^0$ with respect to velocity on the grid, $n_e^0$ is obtained. Having $n_e^0$ and $n_i^0$, one can solve Poisson equation by the well-known fast Fourier transformation (FFT) to obtain $\phi_g^0$ (and therefore $E_g^0=-\partial \phi_g^0/\partial x$). For this purpose, periodic boundary condition is assumed. $\phi_g^0$ can now be exploited in the fluid equations [Eqs. (\ref{3}) and (\ref{4})], while, $E_g^0$ should be interpolated (by a third order Lagrange polynomial interpolation scheme \cite{Numericalrecipes}) to the position of phase points to obtain $E_p^0$. At this stage, we have 
\begin{equation}\label{9}
x_p^0,~~~v_p^0,~~~E_p^0,~~~n_i^0,~~~v_i^0,~~~\phi_g^0. 
\end{equation}
 
Our goal is to find the above quantities at $t=\Delta t$ ($n=1$). To do that, the Leapfrog scheme is employed for Eqs. (\ref{3}), (\ref{4}), (\ref{7}), and (\ref{8}) as follows
\begin{widetext}
\begin{eqnarray}
&&\left(n_i^1\right)_j = \left(n_i^0\right)_j - \frac{\Delta t}{2\Delta x}\left[\left(n_i^{1/2}v_i^{1/2}\right)_{j+1}-\left(n_i^{1/2}v_i^{1/2}\right)_{j-1}\right], \label{10}\\
&&\left(v_i^1\right)_j = \left(v_i^0\right)_j - \frac{\Delta t}{2\Delta x}\left[\frac12\left(v_i^{1/2}\right)_{j+1}^2-\frac12\left(v_i^{1/2}\right)_{j-1}^2+\left(\phi_g^{1/2}\right)_{j+1}-\left(\phi_g^{1/2}\right)_{j-1}\right], \label{11}\\
&&x_p^1 = x_p^0 + \Delta t ~v_p^{1/2}, \label{12}\\
&&v_p^1 = v_p^0 - \frac{\Delta t}{\alpha}~ E_p^{1/2}, \label{13}
\end{eqnarray}  
\end{widetext}
where the subscript ``$j$" means the quantity at the position $x_j=j \Delta x$. Obviously, we need $x_p^{1/2}$, $v_p^{1/2}$, $E_p^{1/2}$, $n_i^{1/2}$, $v_i^{1/2}$, and $\phi_g^{1/2}$. It turns out, however, that overall accuracy of the Leapfrog method is a very sensitive function of the accuracy of the half-time step quantities. In order to minimize the total errors, the half-time step quantities are computed using a predictor-corrector (Euler-Trapezoidal) method. For this purpose, first, we have to determine all quantities by the Euler method (predictor part) at $t=\Delta t/2$,
\begin{widetext}
\begin{eqnarray}
&&\left(n_i^{1/2}\right)_j = \left(n_i^0\right)_j - \frac{\Delta t}{4\Delta x}\left[\left(n_i^{0}v_i^{0}\right)_{j+1}-\left(n_i^{0}v_i^{0}\right)_{j-1}\right], \label{14}\\
&&\left(v_i^{1/2}\right)_j = \left(v_i^0\right)_j - \frac{\Delta t}{4\Delta x}\left[\frac12\left(v_i^{0}\right)_{j+1}^2-\frac12\left(v_i^{0}\right)_{j-1}^2+\left(\phi_g^{0}\right)_{j+1}-\left(\phi_g^{0}\right)_{j-1}\right], \label{15}\\
&&x_p^{1/2} = x_p^0 + \frac{\Delta t}{2}~v_p^{0}, \label{16}\\
&&v_p^{1/2} = v_p^0 - \frac{\Delta t}{2\alpha}~ E_p^{0}, \label{17}
\end{eqnarray}
\end{widetext}
Now, we have the following quantities with first order of accuracy with respect to $\Delta t$, 
\begin{equation}\label{18}
x_p^{1/2},~~~v_p^{1/2},~~~n_i^{1/2},~~~v_i^{1/2}. 
\end{equation}     

Then, upon interpolating $f_p(x_p^{1/2}, v_p^{1/2})$,  $f_g^{1/2}$ and therefore $n_e^{1/2}$ are obtained. With the Poisson solver $\phi_g^{1/2}$ (and therefore $E_g^{1/2}$) can be calculated from $n_e^{1/2}$ and $n_i^{1/2}$. Finally, $E_g^{1/2}$ should be interpolated to the position of phase points to obtain $E_p^{1/2}$.   

The corrector scheme might be built by integrating Eqs. (\ref{3}), (\ref{4}), (\ref{7}), and (\ref{8}) using the Trapezoidal integration scheme which its accuracy is second order with respect to $\Delta t$. The results is as follows,
\begin{widetext}
\begin{eqnarray}
&&\left(n_i^{1/2}\right)_j = \left(n_i^0\right)_j -\frac12  \left[\frac{\left(n_i^{0}v_i^{0}\right)_{j+1}-\left(n_i^{0}v_i^{0}\right)_{j-1}}{2\Delta x}+\frac{\left(n_i^{1/2}v_i^{1/2}\right)_{j+1}-\left(n_i^{1/2}v_i^{1/2}\right)_{j-1}}{2\Delta x}\right]\frac{\Delta t}{2}, \nonumber \\ 
&&~ \label{19}\\
&&\left(v_i^{1/2}\right)_j = \left(v_i^0\right)_j - \frac12 \left[\frac12\frac{\left(v_i^{0}\right)_{j+1}^2-\left(v_i^{0}\right)_{j-1}^2}{2\Delta x}+\frac12 \frac{\left(v_i^{1/2}\right)_{j+1}^2-\left(v_i^{1/2}\right)_{j-1}^2}{2\Delta x}\right. \nonumber\\
&&~~~~~~~~~~~~~~~~~~~~~~~~~~~~~~~~~~~~~~~\left.+\frac{\left(\phi_g^{0}\right)_{j+1}-\left(\phi_g^{0}\right)_{j-1}}{2\Delta x}+
\frac{\left(\phi_g^{1/2}\right)_{j+1}-\left(\phi_g^{1/2}\right)_{j-1}}{2\Delta x}
\right]\frac{\Delta t}{2}, \nonumber \\ 
&&~ \label{20}\\
&&x_p^{1/2} = x_p^0 + \frac{v_p^{0}+v_p^{1/2}}{2} \frac{\Delta t}{2}, \label{21}\\
&&v_p^{1/2} = v_p^0 - \frac{1}{\alpha}\frac{E_p^{0}+E_p^{1/2}}{2}\frac{\Delta t}{2}, \label{22}
\end{eqnarray}
\end{widetext}
The corrector part is performed in an iterative loop to decrease the Euler error up to a favorite value ($\epsilon$). Thus, all improved quantities after the Trapezoidal steps [Eqs. (\ref{19})-(\ref{22})] of table I are compared with their previous corresponding values and the differences can be iteratively reduced. A typical relative error of this kind for the ion velocity versus number of iterations is sketched in Fig. 3. 
\begin{figure}
\includegraphics[height=7cm,width=9cm]{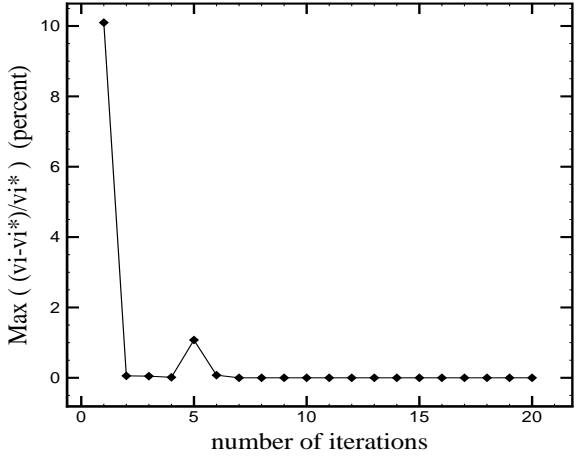} \\
\caption{\label{fig3}The result of employing the Euler-Trapezoidal scheme. The typical relative error in the ion velocity calculation versus number of iterations.}
\end{figure}

This figure illustrates that performing corrector part is quite worthwhile. It is clear that $10$ iterations are often enough for the relative difference of the order of $10^{-6}$. In our case $20$ iterations have been used. Since, the initial loop is used only once in the code, the number of iterations is not a matter. 

The outline of the procedure is given in Table I.
\begin{widetext}
\begin{center}
\begin{tabular}{|l|}
\hline
\begin{tabular}{l}
TABEL I. The initial loop of the code
\end{tabular}
\\ \hline
\begin{tabular}{l}
Initially we have: $x_p^0$, $v_p^0$, $f_p(x_p^0,v_p^0)$, $n_i^0$, $v_i^0$\\ 
\quad  1. Interpolate $f_p(x_p^0,v_p^0)$ to obtain $f_g^0$ and $n_e^0$. \\
\quad  2. Solve Poisson equation to obtain $\phi_g^0$ and $E_g^0$ from $n_e^0$ and $n_i^0$. \\
\quad  3. Interpolate $E_g^0$ to obtain $E_p^0$. \\
\quad  4. The Euler step one: determine $n_i^{1/2}$, $v_i^{1/2}$ [Eqs. (\ref{14}) and (\ref{15})]. \\
\quad  5. The Euler step two: determine $x_p^{1/2}$, $v_p^{1/2}$ [Eqs. (\ref{16}) and (\ref{17})]. \\
\quad  6. Interpolate $f_p(x_p^{1/2},v_p^{1/2})$ to obtain $f_g^{1/2}$ and $n_e^{1/2}$. \\
\quad  7. Solve Poisson equation to obtain $\phi_g^{1/2}$ and $E_g^{1/2}$ from $n_e^{1/2}$ and $n_i^{1/2}$. \\
\quad  8. Interpolate $E_g^{1/2}$ to obtain $E_p^{1/2}$. \\
\quad  9. Trapezoidal step one: determine the ``improved" $n_i^{* 1/2}$, $v_i^{* 1/2}$ [Eqs. (\ref{19}) and (\ref{20})]. \\
\quad 10. Trapezoidal step two: determine the ``improved" $x_p^{* 1/2}$, $v_p^{* 1/2}$ [Eqs. (\ref{21}) and (\ref{22})]. \\
\quad 11. If $\left|n_i^{* 1/2}-n_i^{1/2}\right| > \epsilon $, $\left|v_i^{* 1/2}-v_i^{1/2}\right| > \epsilon $, $\left|x_p^{* 1/2}-x_p^{1/2}\right| > \epsilon $, and $\left|v_p^{* 1/2}-v_p^{1/2}\right| > \epsilon $\\
\quad \quad \quad Then, $n_i^{1/2}=n_i^{*1/2}$, $v_i^{1/2}=v_i^{* 1/2}$, $x_p^{1/2}=x_p^{* 1/2}$, $v_p^{ 1/2} = v_p^{* 1/2}$, and go to 6. \\
\quad \quad \quad Otherwise, pass $n_i^{1/2}$, $v_i^{1/2}$, $x_p^{1/2}$, and $v_p^{1/2}$ to the Leapfrog loop.
\end{tabular}
\\ \hline
\end{tabular}
\end{center}
\end{widetext}
\subsection{The main loop of the code}

So far, we have defined the initial conditions and properly calculated the half-time step of the Leapfrog scheme. Another peculiarity of the Leapfrog scheme that has to be noted is related to the two uncoupled grids defined in the Leapfrog scheme that might cause the two grids drift apart \cite{Potter}. In order to avoid such a decoupling of the grids, we proceed as follows.

Let us first, push the phase point velocities, $v_p^n$, one $\Delta t$,
\begin{equation}
	v_p^{n+1} = v_p^n - \frac{\Delta t}{\alpha}~ E_p^{n+1/2}. \label{23}
\end{equation}
Then, push the phase point positions, $x_p^{n+1/2}$, one $\Delta t$,
\begin{equation}
	x_p^{n+3/2} = x_p^{n+1/2} + \Delta t ~v_p^{n+1}. \label{24}
\end{equation}
Now, we couple the two grids. Calculating $f_g^{*n+3/2}$ by the interpolation of $f_p^*=f_p(x_p^{n+3/2},v_p^{n+1})$ (from hereon, the asterisk superscripts denote the temporary quantities).  The temporary quantities would be corrected in the next steps.
Next, the electron density, $n_e^{* n+3/2}$, is computed by integrating with respect to velocity,
\begin{equation}
	n_e^{* n+3/2} = \int f_g^{*} dv. \label{25}
\end{equation} 
Now, the ion velocity, $v_i^n$, is advanced one time step,
\begin{widetext}
\begin{equation}
\left(v_i^{n+1}\right)_j = \left(v_i^n\right)_j - \frac{\Delta t}{2\Delta x}\left[\frac12\left(v_i^{n+1/2}\right)_{j+1}^2-\frac12\left(v_i^{n+1/2}\right)_{j-1}^2+\left(\phi_g^{n+1/2}\right)_{j+1}-\left(\phi_g^{n+1/2}\right)_{j-1}\right]. \label{26}
\end{equation}
\end{widetext}
Then, the ion density, $n_i^{n+1/2}$ is advanced one time step (coupling of the two grids again),
\begin{widetext}
\begin{equation}
\left(n_i^{* n+3/2}\right)_j = \left(n_i^{n+1/2}\right)_j - \frac{\Delta t}{2\Delta x}\left[\left(n_i^{n+1/2}v_i^{n+1}\right)_{j+1}-\left(n_i^{n+1/2}v_i^{n+1}\right)_{j-1}\right], \label{27}	
\end{equation}
\end{widetext}
The $n_i^{n+1}$ is the average of $n_i^{n+1/2}$ and $n_i^{* n+3/2}$, that is,
\begin{equation}
n_i^{n+1}=\frac12 (n_i^{n+1/2} + n_i^{* n+3/2}). \label{28}	
\end{equation}

Having $n_i^{n+1}$, one can correct $n_i^{* n+3/2}$,
\begin{widetext}
\begin{equation} \label{29}
\left(n_i^{n+3/2}\right)_j = \left(n_i^{n+1/2}\right)_j - \frac{\Delta t}{2\Delta x}\left[\left(n_i^{n+1}v_i^{n+1}\right)_{j+1}-\left(n_i^{n+1}v_i^{n+1}\right)_{j-1}\right].	
\end{equation}
\end{widetext}
In order to continue, Poisson equation should be solved with $n_e^{* n+3/2}$ and $n_i^{n+3/2}$ that leads to $\phi_g^{* n+3/2}$ and after the interpolation to $E_p^{* n+3/2}$. 
 
Now, we push the phase point velocities, $v_p^{n+1}$, one time-step,
\begin{equation}
	v_p^{* n+2} = v_p^{n+1} - \frac{\Delta t}{\alpha}~ E_p^{* n+3/2}. \label{30}
\end{equation}
The corrected $v_p^{n+3/2}$ is the average of $v_p^{n+1}$ and $v_p^{* n+2}$, that is,
\begin{equation}
v_p^{n+3/2}=\frac12 (v_p^{n+1} + v_p^{* n+2}). \label{31}	
\end{equation}

Interpolation of $f_p(x_p^{n+3/2},v_p^{n+3/2})$ leads to the corrected $f_g^{n+3/2}$ and therefore $n_e^{n+3/2}$. Having $n_e^{n+3/2}$ and $n_i^{n+3/2}$, Poisson equation can be solved to obtain $\phi_g^{n+3/2}$ and $E_p^{n+3/2}$ (after the interpolation).

Now, the ion velocity, $v_i^{n+1}$, is advanced one time step (coupling of the two grids one more time),
\begin{widetext}
\begin{equation}
\left(v_i^{* n+2}\right)_j = \left(v_i^{n+1}\right)_j - \frac{\Delta t}{2\Delta x}\left[\frac12\left(v_i^{n+1}\right)_{j+1}^2-\frac12\left(v_i^{n+1}\right)_{j-1}^2+\left(\phi_g^{n+3/2}\right)_{j+1}-\left(\phi_g^{n+3/2}\right)_{j-1}\right]. \label{32}
\end{equation}
\end{widetext}
The corrected $v_i^{n+3/2}$ is the average of $v_i^{n+1}$ and $v_i^{* n+2}$, that is,
\begin{equation}
v_i^{n+3/2}=\frac12 (v_i^{n+1} + v_i^{* n+2}). \label{33}	
\end{equation}

The outline of the algorithm of the main loop is given in Table II.
\begin{widetext}
\begin{center}
\begin{tabular}{|l|}
\hline
\begin{tabular}{l}
TABEL II. The main loop of the code
\end{tabular}
\\ \hline
\begin{tabular}{l}
Initially we just need: $v_p^n$, $x_p^{n+1/2}$, $v_i^n$, $v_i^{n+1/2}$, $n_i^{n+1/2}$, $\phi_g^{n+1/2}$, $E_p^{n+1/2}$\\ 
\quad  1. Push the phase point velocities, $v_p^n$, one $\Delta t$, to obtain $v_p^{n+1}$ [Eq. (\ref{23})]. \\
\quad  2. Push the phase point positions, $x_p^{n+1/2}$, one $\Delta t$, to obtain $x_p^{n+3/2}$ [Eq. (\ref{24})]. \\
\quad  3. Calculate $f_g^{*}$ corresponding to $f_p^*=f_p(x_p^{n+3/2},v_p^{n+1})$ \\
\quad  4. Calculate $n_e^{*n+3/2}$ [Eq. (\ref{25})]. \\
\quad  5. Advance $v_i^n$, one $\Delta t$, to obtain $v_i^{n+1}$ [Eq. (\ref{26})]. \\
\quad  6. Advance $n_i^{n+1/2}$, one $\Delta t$, to obtain $n_i^{* n+3/2}$ [Eq. (\ref{27})]. \\
\quad  7. Determine $n_i^{n+1}$ [Eq. (\ref{28})]. \\ 
\quad  8. Determine $n_i^{n+3/2}$ [Eq. (\ref{29})]. \\ 
\quad  9. Solve Poisson equation with $n_e^{*n+3/2}$ and $n_i^{n+3/2}$ to obtain $\phi_g^{* n+3/2}$ and $E_p^{* n+3/2}$. \\ 
\quad  10. Push the phase point velocities, $v_p^{n+1}$, one $\Delta t$, to obtain $v_p^{* n+2}$ [Eq. (\ref{30})]. \\
\quad  11. Determine  $v_p^{n+3/2}$ [Eq. (\ref{31})]. \\ 
\quad  12. Interpolate $f_p(x_p^{n+3/2},v_p^{n+3/2})$ to obtain $f_g^{n+3/2}$. \\
\quad  13. Calculate $n_e^{n+3/2}$. \\
\quad  14. Solve Poisson equation with $n_e^{n+3/2}$ and $n_i^{n+3/2}$ to obtain $\phi_g^{n+3/2}$ and $E_p^{n+3/2}$. \\
\quad  15. Advance $v_i^{n+1}$, one $\Delta t$, to obtain $v_i^{*n+2}$ [Eq. (\ref{32})]. \\ 
\quad  16. Determine $v_i^{n+3/2}$ [Eq. (\ref{33})]. \\ 
\quad  17. Pass $v_p^{n+1}$, $x_p^{n+3/2}$, $v_i^{n+1}$, $v_i^{n+3/2}$, $n_i^{n+3/2}$, $\phi_g^{n+3/2}$, $E_p^{n+3/2}$ to the next step.\\ 
\end{tabular}
\\ \hline
\end{tabular}
\end{center}
\end{widetext}

\section{Test of the model}

In order to to test the model, we examine the hybrid code both in stationary and non-stationary stages. let us first construct the stationary IA solution of Eqs. (\ref{1})-(\ref{5}). The stationary stage of Eqs. (\ref{3}) and (\ref{4}) are as follows,
\begin{eqnarray}
&&-u_0 \frac{d }{\partial \xi} n_i+ \frac{d}{d \xi}\left(n_i
v_i\right)=0 ,  \label{34} \\
&&-u_0 \frac{d }{d \xi}v_i + v_i \frac {d}{d \xi} v_i= - \frac{%
d}{d \xi}\phi ,  \label{35}
\end{eqnarray}
where $\xi=x-u_0t$ and $u_0$ is the soliton velocity. 

Integrating Eqs. (\ref{34}) and (\ref{35}) and taking into account the necessary conditions for the localized profiles as $\xi \rightarrow \infty$ 
\begin{equation}\label{Boundary}
n_{e,i}\rightarrow 1,~~~~ \phi \rightarrow 0,~~~~ d\phi /d\xi \rightarrow 0,~~~~v_{i}\rightarrow 0.
\end{equation}
Thus, we obtain,
\begin{eqnarray}
&&n_i=\left( 1-\frac{2\phi }{u_0^{2}}\right) ^{-1/2},  \label{36}  \\ 
&&v_i=u_0-\sqrt{u_0^2-2\phi}~. \label{37}
\end{eqnarray}  

The electron density is obtained from Eq. (\ref{2}). Therefore, the stationary solution of Eq. (\ref{1}) has to be introduced. Based on the polarity of soliton when $-\phi$ is a potential well, a number of electrons might be in resonance with it and through a nonlinear mechanism, are trapped. The model distribution function, containing both the free and trapped electrons in the Maxwellian plasma was first introduced by \cite{Schamel}. It is a distorted Maxwellian that has a hole-like structure near the IA soliton velocity, $u_0$, as follows,
\begin{widetext}
\begin{eqnarray}
&&f_{f}=\left\{
\begin{array}{l}
\sqrt{\alpha/(2\pi)}\exp \left[-\frac12 \left(\sqrt{\alpha}u_0-\sqrt{2\epsilon_e}\right)^2\right],\qquad v_e<u_{0}-\sqrt{2\phi/\alpha } \\
\sqrt{\alpha/(2\pi)}\exp \left[-\frac12 \left(\sqrt{\alpha}u_0+\sqrt{2\epsilon_e}\right)^2\right],\qquad v_e>u_{0}+\sqrt{2\phi/\alpha }
\end{array}
\right. \label{39}\\
&&f_t=\sqrt{\alpha/(2\pi)}\exp \left(-\frac12 \alpha u_0^2 - \beta \epsilon_e\right),~\qquad u_{0}-\sqrt{2\phi/\alpha}\leq v_e \leq u_{0}+\sqrt{2\phi/\alpha}\label{40a}
\end{eqnarray}
\end{widetext}
where $f_{f}$ and $f_{t}$ are the free and trapped parts of electron distribution function, respectively, and 
\begin{equation}\label{40} 
\epsilon_e=\frac12\alpha\left(v_e-u_0\right)^2-\phi.
\end{equation}
Then, the electron density is obtained by integrating the distribution functions over the corresponding velocity range. 

Having $n_e$ and $n_i$, the last step for the stationary IA soliton is the solution of Poisson equation,
\begin{equation}\label{41}
\frac{d^2 \phi}{d \xi^2}=n_e-n_i.
\end{equation}

The localized solution of Eq. (\ref{41}) is numerically obtained and shown in Fig. 4.
\begin{figure}
\includegraphics[height=7cm,width=9cm]{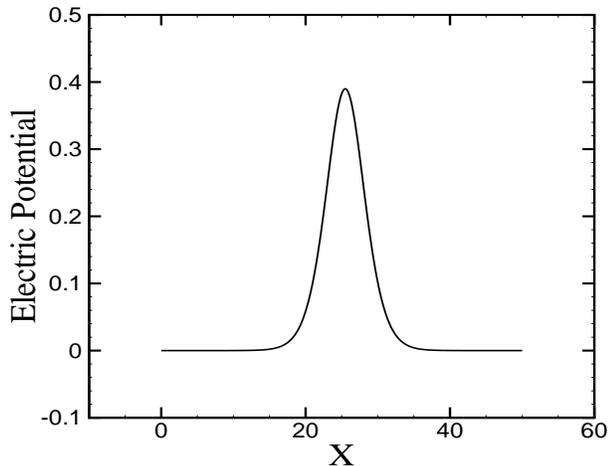} \\
\caption{\label{fig4}The stationary IA soliton that is used as the simulation test.}
\end{figure}
Equation (\ref{41}) is a nonlinear boundary value problem. Discretizing Eq. (\ref{41}) in the configuration space, we obtain
\begin{equation}\label{Newton}
\phi_{j+1}-2 \phi_{j}+\phi_{j-1}+(\Delta x)^2\left[\left(n_e\right)_j-\left(n_i\right)_j\right]=0. 
\end{equation} 
Thus, the obtained potential $\phi$ is second order accurate in $\Delta \xi (=\Delta x)$. In this way, we have transformed Eq. (\ref{41}) to a set of $Nx$ simultaneous nonlinear equations (there are $Nx$ grid points along the $x$ axis). Recall that the boundary condition was introduced through Eq. (\ref{Boundary}). The set of equations have been solved by the Newton iterative method \cite{Newton}. Each iteration deals with a matrix equation (containing a tridiagonal matrix) that has been solved by the recurrence method \cite{Potter}. 

Now it's time to set up the non-stationary experiment, i.e. disintegration of an initial Gaussian profile into an IA soliton train. To this end, the  initial potential $\phi$ is defined as,
\begin{equation}\label{Gaussian}
\phi = A\exp \left[ -\left(\frac{x-C}{\Delta}\right)^2\right],
\end{equation}   
where $A$ is the amplitude, $\Delta$ is the half of the width and $C$ is introduced to fix the position of the maximum of the initial Gaussian profile in the simulation box. In order to present a variety of test problems, the polarity of the potential is chosen in such a way that  electrons feel an obstacle. Therefore, the initial electron velocity distribution is considered as follows,
\begin{equation}\label{distribution}
f={\sqrt \frac{\alpha}{2\pi}}\exp\left(\frac12 \alpha v^2+\phi\right).
\end{equation}
Accordingly, the initial electron density is
\begin{equation}\label{electron}
n_{e}=\exp\left(\phi\right).
\end{equation}
Then, the initial ion density can be obtained by Poisson equation in the following form,
\begin{equation}\label{ion}
n_{i} = n_{e} - \frac{d^2 \phi}{dx^2}.
\end{equation}
In order to define the appropriate initial ion velocity we proceed as follows. Since the initial profile is supposed to be disintegrated into several solitons through a slowly varying dynamics on the time scale associated with $\omega_{pi}$ and for each solitons Eqs. (\ref{36}) and (\ref{37}) are necessary, an appropriate candidate for the initial ion velocity might be
\begin{equation}\label{velocity}
v_{i} = \sqrt{\frac{2 \phi (n_{i}-1)}{n_{i}+1}},
\end{equation}
that is obtained after substituting $u_0$ from Eq. (\ref{36}) in Eq. (\ref{37}). Although, Eq. (\ref{velocity}) is held in the stationary state for solitonic potential, our insight about the slowly varying dynamics led us to deduce that if the initial condition fulfills Eq. (\ref{velocity}), the non-stationary evolution of the Gaussian profile will be given rise to a soliton train. 

\section{The experiment}
  
In this section, the hybrid code is examined by the test problems introduced in the previous section. First is the propagation of a stationary IA soliton. For this purpose, assume $\beta=-0.5$ and $u_0=1.5$. Then, $f_p$ is built by substituting $\phi$ (the numerical solution of Eq. [\ref{41})] in Eqs. (\ref{39}) and (\ref{40a}). Moreover, the initial ion density and velocity are constructed by substituting $\phi$ in Eqs. (\ref{36}) and (\ref{37}), respectively. Therefore, to complete  the initial conditions, we just need to set the phase points in the phase space (look at the first row of Table I).  In this experiment the following parameters are considered,
\begin{eqnarray}\label{42}
&&Lx=50,~~~~ \Delta x = 0.05,~~~~ \Delta v_e = 0.1,~~~~ \nonumber \\
&&\Delta t = 0.01,~~~~ v_{e,max}=300,~~ v_{e,min}=-300.
\end{eqnarray}
where, ``$Lx$'' is the total length of the configuration space. It is clear that the total length of the velocity space is $600$. Due to computational constraints, velocity cutoffs as in PIC models are imposed ($-300<v_e<300$) \cite{2}. Moreover, there are $9$ phase points in each cell of the phase space, Fig. 1.   

\begin{figure}
\includegraphics[height=7cm,width=9cm]{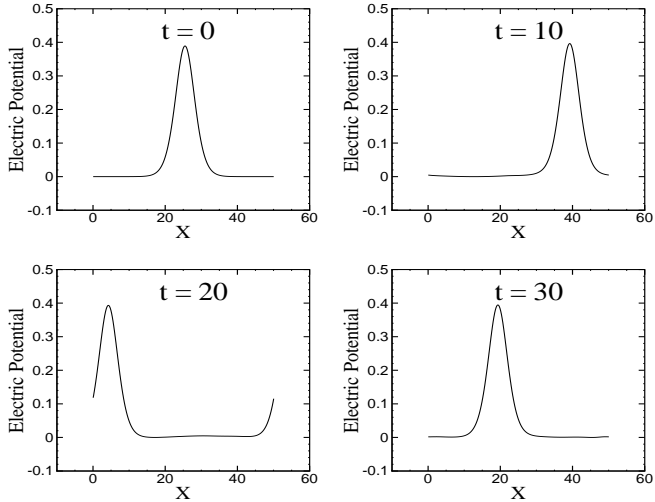} \\
\caption{\label{fig5}The propagation of the IA soliton from $t=0$ to $t=30$.}
\end{figure}

Figure 5 depicts the results associated with the soliton propagation from $t=0$ to $t=30$. As it is seen in Fig. 5, the electric potential, $\phi$, moves in the simulation box while its shape remains unchanged. Since $\beta$ is negative, the electron distribution function contains a hole in the phase space. 
Fig. 6 shows the electron distribution function with the mentioned hole structure. It is obvious that it moves along the configuration axis with constant velocity. Note that the velocity of the hole is constant since there isn't any displacement along the velocity axis. 
\begin{figure}
\includegraphics[height=7cm,width=8cm]{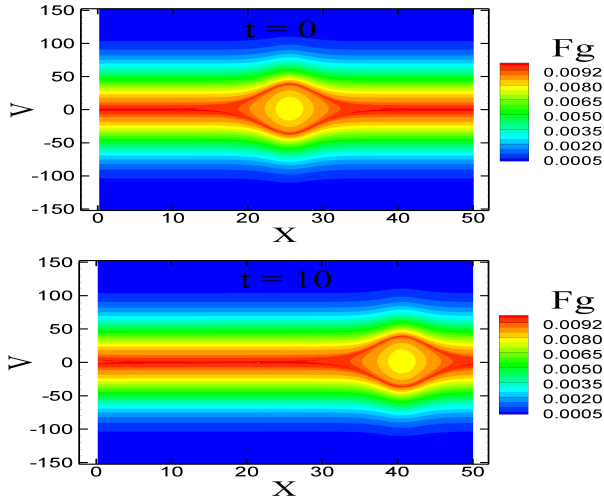} \\
\caption{\label{fig6}The electron distribution function in the phase space. The electron hole, associated with the IA soliton, moves toward the right side.}
\end{figure}
A space-temporal evolution diagram of the electric potential, in contour form, is shown in Fig. 7. In the case of soliton with constant velocity, the maximum of the electric potential has to lie on a straight line of the space-temporal plot that its slope is equal to soliton speed. The figure confirms the constancy of the soliton velocity. The slope of the lines of Fig. 7 has been calculated and is $1.499$. Besides, in the figure, the periodic boundary along the ``$x$'' axis is realizable as the repeated structures.  

\begin{figure}
\includegraphics[height=7cm,width=9cm]{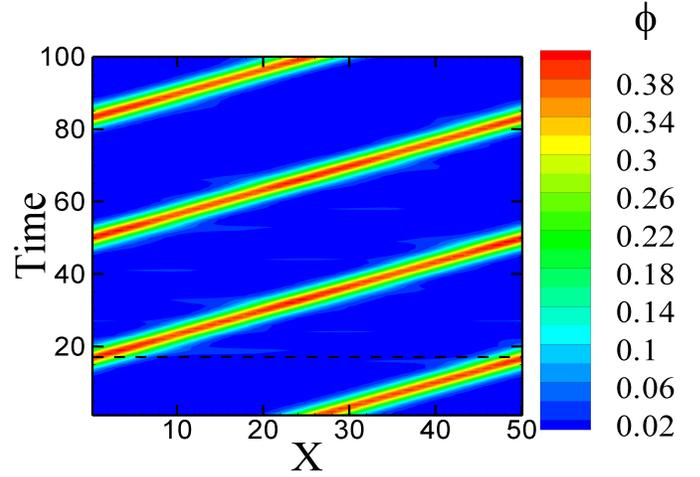} \\
\caption{\label{fig7}The space-temporal evolution diagram of the electric potential. The straight line is an indication that the soliton moves with constant velocity.}
\end{figure}
Conservation laws are the other tests that are used to demonstrate the accuracy of the simulation model. The most basic tests are the energy and entropy conservations. The model is collisionless and therefore both the total energy (the energy of field and particles) and the entropy are constants of motion. Figure 8 exhibits the relative error in the total energy [i.e. (total energy$^{n}$-total energy$^0$)/total energy$^0$ $\times 100$, recalling that the superscript ''$n$" denotes the quantities at  $t=n \Delta t$]. 
\begin{figure}
\includegraphics[height=5cm,width=9cm]{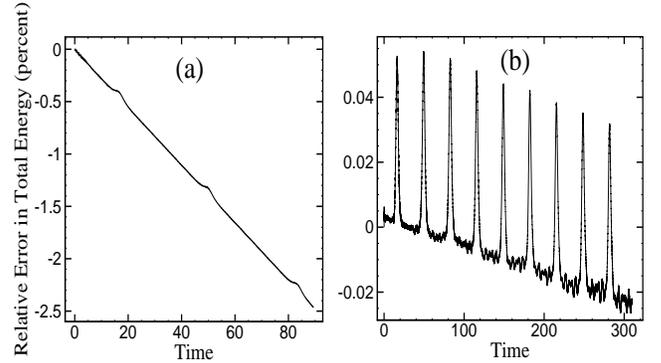} \\
\caption{\label{fig8}The relative error in the total energy. a) The energy conservation when the first step of the Leapfrog scheme is obtained by the Euler method. b) The energy conservation when the first step of the Leapfrog scheme is obtained by the Euler-Trapezoidal. The periodic behavior is due to the periodic boundary condition.}
\end{figure}
As it was mentioned, the Euler scheme is not enough accurate to be used, as the initial half-time step, by the Leapfrog scheme. Figure 8a demonstrates this fact as the result of the hybrid code execution. As it seen, at $t=100$ there is almost three percent error. Figure 8b shows the result of applying the Euler-Trapezoidal scheme (see Table I) as the initial loop. It is clear that the error has been remarkably reduced. This justifies our insistence on introducing the Euler-Trapezoidal scheme for the initial loop. 
Figure 9 depicts the relative error of the entropy (that is calculated in the same manner as was used in the energy case). It is clear that during the execution time, the relative error of the entropy is limited to maximum $0.02\%$.

\begin{figure}
\includegraphics[height=7cm,width=9cm]{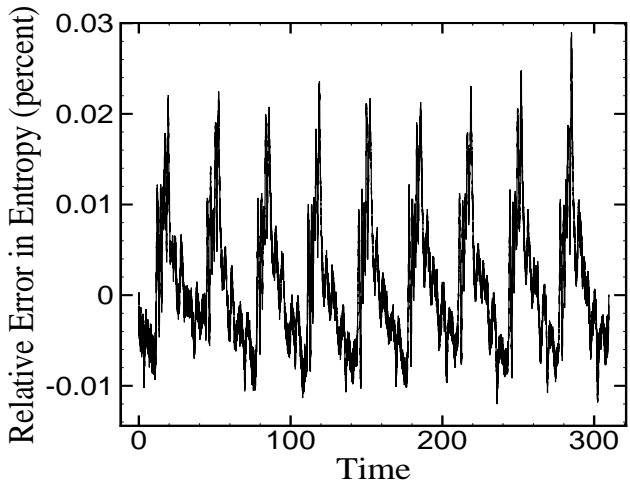} \\
\caption{\label{fig9}The relative error in the entropy. Whenever, the soliton passes the boundary, a jump takes place.}
\end{figure}
With a glance in Figs. 8 and 9, a kind of periodic behavior is distinguishable. It is an  important effect that is related to the periodic boundary condition. There are different way of constructing periodic boundary condition (connecting the beginning and the ending of the configuration space). The influence of using periodic boundary condition exhibits itself as an inhomogeneity in the configuration space. Therefore, whenever the soliton passes the boundary, a small part of it would be reflected. The reflected part is very small, however, it would be a problem for long-time execution. That is, passing the boundary causes a jump in the error (both in the energy and entropy). To avoid this problem, another version of the code has been designed in which all the velocities has been transferred to the measure of the IA velocity (Galilean Transformation). Therefore, in this version the soliton have to be immobile and wouldn't pass the boundaries. The results is shown in Figure 10. The figure is another indication that the soliton velocity is exactly $u_0=1.5$ and it moves without any change. Since, after elapsing of 100 unit of time, there is not any considerable difference in comparison to the initial soliton. Moreover, it is obvious from Fig. 11, that the relative error in the total energy has been considerably reduced and the periodic behavior is not seen anymore.  
\begin{figure}
\includegraphics[height=7cm,width=9cm]{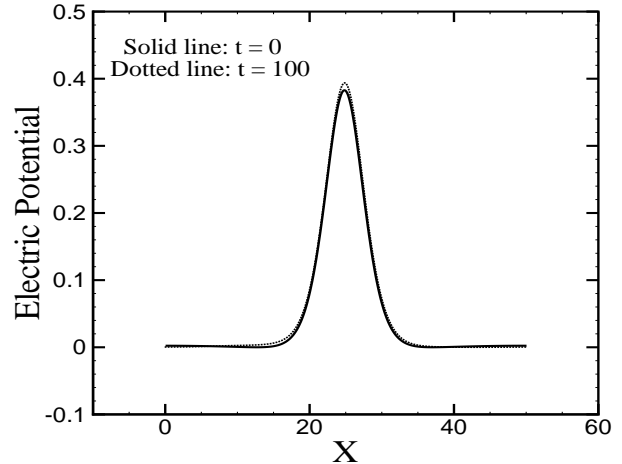} \\
\caption{\label{fig10}The immobile IA soliton as the result of executing the hybrid code with moving grid.}
\end{figure}
\begin{figure}
\includegraphics[height=7cm,width=9cm]{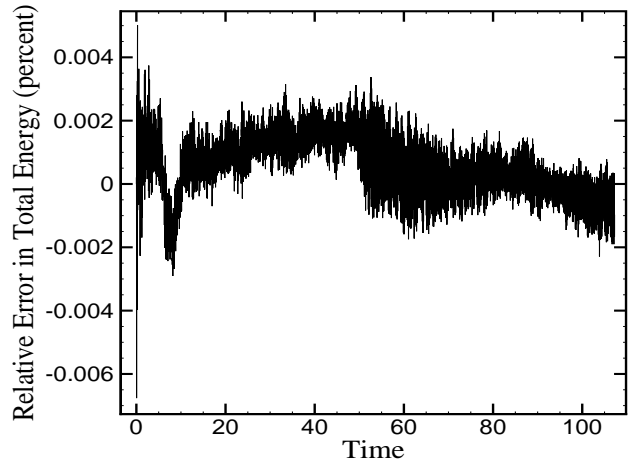} \\
\caption{\label{fig11}The relative error in the total energy of the simulation in moving grid. The reduction of error, in comparison to Fig. 8, is obvious}
\end{figure}

The second experiment is devoted to disintegration of a Gaussian profile into IA soliton train. Figure 12 shows the result of the evolution of ion density for $A=0.2$, $\Delta=20$, and $C=64$ at $T=900$. The disintegration of the initial Gaussian profile leads into three IA solitons and a linear IA wave in the back of the initial profile. The dotted straight line that is used to connect the maximums is an indication of well-known fact that IA solitons velocity (in the absence of trapped electrons) is directly proportional to their amplitudes.   
\begin{figure}
\includegraphics[height=7cm,width=9cm]{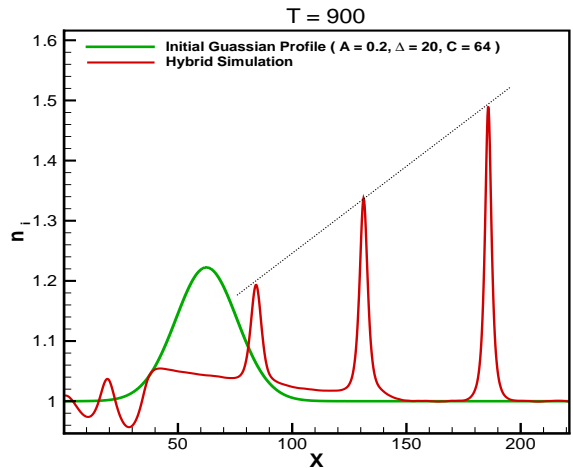} \\
\caption{\label{fig12}Non-stationary experiment of the Hybrid code. The disintegration of an initial Gaussian profile into three IA solitons and a linear IA wave.}
\end{figure}

\section{conclusion}
A hybrid scheme was introduced for simulating the coupled Vlasov-fluid-Poisson system. It was designed for a collisionless plasma with hot electrons and cold positive ions. The Vlasov equation was solved for the electrons and the ions followed the fluid equations. The periodic boundary conditions were assumed. The method of solution of the collisionless Vlasov equation was based on following fixed collisionless phase point trajectories. It was done by solving the characteristic equations of the Vlasov equation. Using the average interpolation scheme in phase space,  the electron distribution function was mapped to a fixed background phase space grid while retaining it at the phase point. Both, the characteristics equation and fluid equations were solved using the Leapfrog method. However, to obtain the first half-time step of the Leapfrog, the Euler-Trapezoidal scheme was introduced. The presented scheme conveniently coupled the two well-known grids in the Leapfrog method. The first test of the model was the propagation of an stationary IA soliton.  The simulation code preserved the stationary soliton features. Conservation laws were the other benchmark tests. The error in the relative entropy and total energy was kept to less than one percent. As the non-stationary test, disintegration of a Gaussian profile into IA solitons was considered and confirmed the appropriate performance of the hybrid code once more.

\end{document}